\documentclass[12pt]{article}   	
\usepackage{geometry,bbm}   
\usepackage[toc,page]{appendix}             		
\usepackage{graphicx}
\usepackage[usenames,dvipsnames]{color}
\usepackage{slashed}					
\usepackage{amssymb,amsmath}
\usepackage{hyperref}
\usepackage{mathtools}
\usepackage{multirow}
\usepackage[table]{xcolor}

\def\){\right)}
\def\({\left( }
\def\]{\right] }
\def\[{\left[ }

\newcommand{\tx}{\text}

\newcommand{\w}{\wedge}

\newcommand{\ea}{\end{align*}}
\newcommand{\ba}{\begin{align*}}
\newcommand{\bi}{\begin{itemize}}
\newcommand{\ei}{\end{itemize}}

\newcommand{\ben}{\begin{enumerate}}
\newcommand{\een}{\end{enumerate}}
\newcommand{\ol}{\overline}

\def\){\right)}
\def\({\left( }
\def\]{\right] }
\def\[{\left[ }
\newcommand{\la}{\langle}
\newcommand{\ra}{\rangle}

\def\NO{\nonumber}

\newcommand{\be}{\begin{equation}}
\newcommand{\ee}{\end{equation}}

\def\bea{\begin{eqnarray}}
\def\eea{\end{eqnarray}}

\def\bal#1\eal{\begin{align}#1\end{align}}

\def\bald{\begin{aligned}}
\def\eald{\end{aligned}}

\def\beqx{\begin{displaymath}}
\def\eeqx{\end{displaymath}}

\newcommand{\bmat}{\left(\begin{array}}
\newcommand{\emat}{\end{array}\right)}

\def\a{\alpha}
\def\b{\beta}

\def\d{\delta}
\def\e{\epsilon}
\def\f{\phi}
\def\g{\gamma}

\def\l{\lambda}
\def\m{\mu}
\def\n{\nu}
\def\o{\omega}
    
\def\p{\pi}

\def\s{\sigma}
\def\t{\tau}

\def\D{\Delta}
\def\F{\Phi}

\def\vf{\varphi}

\def\ba{\bbalpha}

\def\cn{{\cal N}}

\def\cw{{\cal W}}

\def\bo{{\raise-.3ex\hbox{\large$\Box$}}}               
\def\pa{\partial}                                       
\def\face{{\raise.2ex\hbox{$\displaystyle \bigodot$}\mskip-2.2mu \llap {$\ddot
        \smile$}}}                                   
\def\>{\rangle}                                      
\def\<{\langle}                                      

\def\leftrightarrowfill{$\mathsurround=0pt \mathord\leftarrow \mkern-6mu
        \cleaders\hbox{$\mkern-2mu \mathord- \mkern-2mu$}\hfill
        \mkern-6mu \mathord\rightarrow$}        
\def\dvec#1{\vbox{\ialign{##\crcr
        \leftrightarrowfill\crcr\noalign{\kern-1pt\nointerlineskip}
        $\hfil\displaystyle{#1}\hfil$\crcr}}}           
\def\Tr{{\rm Tr \,}}                                    
\def\Im{{\rm Im\,}}                                     

\def\-{\hphantom{-}}

\numberwithin{equation}{section}
\begin{document}

\begin{titlepage}

\begin{flushright}
SISSA 50/2016/FISI
\end{flushright}
\bigskip
\def\thefootnote{\fnsymbol{footnote}}

\begin{center}
{\LARGE
{\bf
Broken current anomalous dimensions, \\ \vskip 20pt conformal manifolds and RG flows }}
\end{center}

\bigskip
\begin{center}
{\large
Vladimir Bashmakov$^{1}$, Matteo Bertolini$^{1,2}$, Himanshu Raj$^1$}

\end{center}

\renewcommand{\thefootnote}{\arabic{footnote}}

\begin{center}
\vspace{0.2cm}
$^1$ {SISSA and INFN - 
Via Bonomea 265; I 34136 Trieste, Italy\\}
$^2$ {ICTP - 
Strada Costiera 11; I 34014 Trieste, Italy\\}

\vskip 5pt
{\texttt{bertmat,hraj,vbashmakov @sissa.it}}

\end{center}

\vskip 5pt
\noindent
\begin{center} {\bf Abstract} \end{center}
\noindent
We consider deformations of a conformal field theory that explicitly break some global symmetries of the theory. If the deformed theory is still a  conformal field theory, one can exploit the constraints put by conformal symmetry to compute broken currents anomalous dimensions. We consider several instances of this scenario, using field theory techniques and also holographic ones, where necessary. Field theoretical methods suffice to discuss examples of symmetry-breaking deformations of the $O(N )$ model in 
$d=4-\e$ dimensions. Holography is instrumental, instead, for computing current anomalous dimensions in $\b$-deformed superconformal field theories, and in a class of supersymmetric RG flows at large $N$.

\vspace{1.6 cm}
\vfill

\end{titlepage}

\newpage
\tableofcontents

\section{Introduction}
\label{Introduction}

Conformal field theories (CFTs) play a central role in physics. Several physical phenomena are governed by (approximate) CFTs and in the theoretical understanding of many others they are key ingredients. In fact, the extreme UV and IR dynamics of a generic quantum field theory (QFT) are often governed by a CFT,\footnote{This is known to be true at least in $d=2$ and $d=4$ dimensions (and believed to be so for $d\leq 6$). In this work, we will be mostly concerned with four-dimensional QFTs.} so the latter are also important to have control on QFT in general.

The basic data one needs to know to characterize a CFT are the spectrum of primary operators and OPE coefficients, and different approaches can be pursued to have control on them (e.g., the recent renewed interest in the bootstrap program \cite{Rattazzi:2008pe}).  

Whenever two conformal field theories are connected by some deformation, be it relevant or marginal, interesting phenomena can arise. For instance, it can happen that a primary operator of the undeformed CFT enjoying some shortening condition merges with another primary, and the two distinct conformal families {\it recombine} into a single, longer one in the deformed CFT  \cite{Rychkov:2015}. This phenomenon can be used to understand properties of the deformed CFT once the original CFT and the corresponding deformation are known. For instance, if the deformation is small,  one can compute the leading correction to the anomalous dimension of operators which recombine just by doing computations in the original CFT  \cite{Anselmi:1998ms}. 

In this paper we will focus on the phenomenon of multiplet recombination, in particular of {\it current} multiplets. The basic dynamics can be summarized as follows. 
Consider a $d$-dimensional CFT with some global symmetry (think of a $U(1)$ symmetry, for definiteness). The corresponding current is conserved
\be
\label{conscur1}
\partial^\m J_\m = 0~.
\ee
From the above equation it follows that the CFT operator $J_\mu$ is at the unitarity bound, its dimension being $\Delta_{J} = d-1$. 

Suppose now we perturb the CFT by a deformation triggered by some charged scalar operator and coupling $g$. Such a deformation breaks the symmetry explicitly. Suppose further that the deformed theory is still a CFT; either because the deformation is (exactly) marginal or, if the deformation is relevant, because the end point of the RG is still a CFT. At such a fixed point the current is not conserved anymore, that is
\be
\label{brokencur1}
\partial^\m J_\m = {O}~,
\ee
with ${O}$  an operator of the deformed CFT, itself related to a scalar primary operator of the undeformed CFT.\footnote{This primary operator is the one obtained by acting with a symmetry transformation on the operator triggering the deformation. We are assuming that for every local operator of the deformed theory there exists one in the undeformed theory such that correlation functions of the former reduce to those of the latter for $g=0$.} Note that in the deformed CFT ${O}$ is a descendant of the spin-one current, so they belong to the same conformal family, while in the original CFT they do not, see eq.~\eqref{conscur1}. 

Strictly speaking, this picture holds only when currents are weakly broken. Namely, when the symmetry-breaking CFT sits at small values of the coupling $g$, anomalous dimensions are small and can be evaluated perturbatively in $g$. This is the case whenever the deformation is marginal, since then $g$ does not run and can be taken to be arbitrarily small. For relevant deformations, instead, the coupling runs and reaches some fixed value $g_*$. In this case, there should exist some parameter which lets one tune $g_*$ to small values; i.e. it should be possible to make the RG flow arbitrarily short in the space of couplings. For generic RG flows, instead, $g_*$ cannot be tuned to zero, anomalous dimensions can easily be order 1, and it is difficult to map conformal families of the two CFTs. Still, it remains true that in the deformed theory there is (at least) a spin-1 short operator less, and a spin-1 long operator more (with all its descendants).\footnote{This is true as long as there are no emergent symmetries in the IR. The latter, however, would not affect the multiplet recombination we are discussing, and hence we do not consider such a possibility.} Therefore, in the following, we will refer to current multiplet recombination even for these more intricate situations. 

From eq.~\eqref{brokencur1} it follows that the scaling dimension of the current in the deformed CFT is $\Delta_{J} > d-1$, meaning that some (positive by unitarity) anomalous dimension has been generated 
\be
\label{anJ1}
\Delta_{J} = d-1 + \g~.
\ee
From the viewpoint of representations of the conformal algebra, the multiplet where $J_\mu$ sits is a short multiplet in the original CFT, and a long multiplet in the deformed one. As seen from the original CFT, the multiplet to which $J_\m$ belongs has recombined with that to which ${O}$ does. 

In this paper, we present several concrete realizations of this scenario, triggered by exactly marginal and relevant deformations, and compute anomalous dimensions of broken currents. 

If the undeformed CFT is a free theory and the deformation can be made parametrically small, it turns out that field theory techniques suffice to reach such a goal. An example we will discuss in some detail is the $O(N)$ model in $d=4-\e$ dimensions, for which relevant, symmetry-breaking deformations can be considered, and IR fixed points are reached for parametrically small values of the couplings. 

For interacting, possibly strongly coupled CFT, instead, we will turn to holography and compute $\g$ using AdS/CFT techniques. This will allow us to compute the anomalous dimension of broken currents in a class of $\cn=1$ superconformal field theories (SCFTs) arising from D-branes at toric Calabi-Yau singularities which admit symmetry-breaking exactly marginal deformations known as $\b$ deformations \cite{Leigh:1995ep,Benvenuti:2005wi}. Holography will also be instrumental in discussing a class of RG flows connecting SCFT at strong coupling. Here the anomalous dimensions will be large, but the possibility to have access to the entire flow allows one to have control on how multiplets recombine and to compute current anomalous dimensions. 

In the next section we explain the field theory and holographic methods we use. We will be interested both in deformations driven by relevant operators and by marginal ones. Therefore, we will also review some basic results regarding the possible existence of exactly marginal deformations in conformal field theories. In section \ref{marsec} we will present examples where the deformation is driven by exactly marginal operators. We will start with a toy-model, which can be described within field theory, and then discuss $\beta$-deformed ${\cal N}=1$ SCFT. In section \ref{relsec} we will consider, instead, instances where the symmetry-breaking deformation is relevant. First we discuss the $O(N)$ model, which can be treated using field theory techniques, and then focus on a class of holographic models describing RG flows between SCFTs at strong coupling \cite{Ceresole:2001wi}. Section \ref{consec} contains our conclusions. Three appendices contain some technical material we did not include in the bulk of the paper.

\section{Methods - Field theory and holography}
\label{secmethod}

Computing operator anomalous dimensions exactly is, in principle, very difficult. However, when these arise because multiplets recombine in a CFT, eq.~\eqref{brokencur1}, the constraints put by conformal symmetry help. 

Let us first suppose that the breaking is weak. This means that the CFT with broken symmetries can be made parametrically near the symmetry-preserving one. To make this manifest, let us rewrite eq.~\eqref{brokencur1} as
\be
\label{brokencur2}
\partial^\m J_\m = g \, O~,
\ee
which is just to emphasize that at $g=0$ the current is conserved. Here we are considering either deformations triggered by exactly marginal operators or by relevant ones. In the latter case $g$ should be understood as $g_*$, the value of the coupling at the IR fixed point, which, as such, is dimensionless. 

In such a situation, as we will review below, one can determine $\g$, to {\it leading} order in $g$, by computing the two-point functions of $O$ and $J_\mu$ in the unperturbed CFT \cite{Anselmi:1998ms,Belitsky:2007jp} (interesting recent works using a similar approach are Refs.~\cite{Rychkov:2015,Basu:2015gpa,Skvortsov:2015pea,Giombi:2016hkj}). The basic idea goes as follows.

In a CFT, the structure of two-point functions of primary operators is fixed, up to an overall normalization, by conformal invariance. In particular, we have for the spin-one current
\bea
\label{2pfj}
\<J_{\mu}(x)J_\nu(y)\> & = &  C_J \, \frac{I_{\m\n}}{(2\pi)^d\(x-y\)^{2 \Delta_J}},~~~~I_{\m\n}=\delta_{\m\n} - 2 \frac{\(x-y\)_\m \(x-y\)_\n}{\(x-y\)^2}~.
\eea
This equation holds independently of the current being or not being conserved, so, in our case, both in the unperturbed CFT and in the perturbed one. However, the operator dimension $\Delta_J$ as well as $C_J$ differ, since they depend on $g$ (in fact, on any coupling of the theory, in general).

Differentiating the correlator \eqref{2pfj} twice, one gets
\be
\label{2pfdj}
\<\partial^\m J_{\m}(x) \partial^\n J_\n(y)\> =  C_J \, \frac{2 (2\Delta_J + 2 -d)  (\Delta_J + 1 - d)}{(2\p)^d(x-y)^{2 \Delta_J + 2}} ~.
\ee
By the operator identity \eqref{brokencur2} the same two-point function is given by 
\be
\label{2pfdj2}
\<\partial^\m J_{\m}(x) \partial^\n J_\n(y)\>= g^2  \< O(x) O(y)\>~.
\ee
By taking the ratio with \eqref{2pfj}, using eq.~\eqref{anJ1}, one gets 
\be
\label{lgCrel1}
g^2 \, (x-y)^{2} I_{\m\n} \frac{\<O(x)O(y)\>}{\<J_{\mu}(x)J_\nu(y)\>}=   2 \g  \( d + 2\g \)~.
\ee
The above equation shows that in computing current anomalous dimension one needs to know the correlators to one order less in perturbation theory. In particular, to get $\g$ to leading order in $g$ one needs the value of the two-point functions of $J_\m$ and of $O$ at zeroth order, namely in the undeformed theory where $O$ is a primary operator (cf the footnote in page two) and its two-point function has the following structure
\bea
\label{2pfO}
\<{O}(x){O}(y)\> &=& C_{{O}}\, \frac{1}{(2\p)^d\(x-y\)^{2 \Delta_{{O}}}}~.
\eea
Plugging this expression into eq.~\eqref{lgCrel1} and using eq.~\eqref{2pfj} one gets, upon expanding in powers of the coupling $g$ (note that eq.~\eqref{brokencur2} implies that $\Delta_{O} = \Delta_J +1$)
\be
\label{lgCrel2}
\g  =  \frac{1}{2d} \, g^2 \, \frac{C_{O}}{C_J}  + O(g^4)~,
\ee
with $C_J$ and $C_{O}$ evaluated in the undeformed theory, namely at $g=0$.

This method is powerful because it allows one to get information on the deformed CFT by just doing computations in the undeformed one. In practice, however, there are two limitations. First, as already emphasized, the perturbative expansion \eqref{lgCrel2} makes sense only if the symmetry is weakly broken. If this is not the case, the above strategy cannot be applied  and one should resort to some other method. Second, computing the two-point functions of $O$ and $J_\m$, and hence the exact proportionality coefficient in eq.~\eqref{lgCrel2}, is straightforward only if the undeformed CFT is a free theory. In such a case one has to deal with correlators at tree level and there are no issues of regularization and renormalization. A different story is if the original CFT is an interacting, possibly strongly coupled theory, e.g., emerging from some non-trivial gauge theory dynamics. These are all situations where AdS/CFT techniques can come to the rescue (for field theories with a holographic dual). 

In AdS/CFT, QFT global currents are dual to gauge fields in the bulk, the mass/dimension relation, in units of the AdS radius, being  
\be
\label{massdimcur1}
m^2 = d-1 + \Delta_J (\Delta_J - d)~.
\ee
From eq.~\eqref{massdimcur1} it follows that massless gauge fields are dual to conserved currents, and massive ones to non-conserved currents. Therefore, when two CFTs are related by a symmetry-breaking deformation the gauge field dual to the (broken) current is massless in the vacuum dual to the undeformed CFT, and massive in that dual to the deformed CFT.  Indeed, as known since the early days of the AdS/CFT correspondence, the breaking of a field theory global symmetry (be it explicit, like in our case, or spontaneous) corresponds to a Higgs mechanism in the bulk, by which a massless vector eats up a scalar and becomes massive. This is the bulk counterpart of the dynamics which governs current multiplet recombination.\footnote{See Ref.~\cite{Bashmakov:2016pcg} for a holographic description of scalar multiplet recombination, and Refs.~\cite{Bianchi:2003wx,Beisert:2004di,Bianchi:2004ww,Bianchi:2005ze,Klebanov:2002ja,Girardello:2002pp,Leigh:2012mz} for that of higher-spin currents. These are both described by a Higgs-like mechanism in the bulk, though of a different nature in the two cases.} Therefore, to compute current anomalous dimensions holographically, one has to calculate the mass of the dual gauge field and plug the result into eq.~\eqref{massdimcur1}. Note that this provides the anomalous dimension at face value so it also applies to long RG flows, i.e. when $g_*$ cannot be tuned to zero. In later sections, we will discuss instances of this kind. Another situation in which AdS/CFT techniques can help is when the breaking is weak but the undeformed CFT is itself at strong coupling, and therefore computing at $g=0$ is itself non-trivial. In this case, one can evaluate the two-point functions $\<J_{\mu}(x)J_\nu(y)\>$ and $\<{O}(x){O}(y)\>$ entering eq.~\eqref{lgCrel1} holographically. The $\b$-deformed SCFTs we will discuss later are one such example. 

\subsection{On exactly marginal deformations}

Current multiplet recombination can be triggered by relevant or by exactly marginal deformations. When the deformation is relevant, an RG flow is induced. When the deformation is marginal, instead, there is no RG flow. One is moving along the conformal manifold, the space of exactly marginal deformations ${\cal M}_c$. 
 
The existence of exactly marginal deformations is difficult to establish, and for a generic CFT they do not exist, in general. However, as shown originally by Leigh and Strassler \cite{Leigh:1995ep}, and further elaborated by, e.g., Refs.~\cite{Kol:2002zt,Benvenuti:2005wi,Green:2010da,Kol:2010ub}, four-dimensional ${\cal N}=1$ SCFTs often enjoy non-trivial conformal manifolds. 

Suppose we have a SCFT with some global symmetry group $G$ and a bunch of marginal chiral operators ${\cal O}_i$ carrying some non-trivial representation of $G$. Deforming the theory by a $G$-breaking marginal superpotential ${\cal W} = \sum_i g^i {\cal O}_i$, an RG flow is induced since, generically, the operators ${\cal O}_i$ acquire an anomalous dimension.\footnote{In a SCFT, there do not exist marginal K\"ahler deformations \cite{Green:2010da}. Therefore,  marginal deformations are described by superpotential deformations.
} In fact, marginal operators may either remain marginal or become marginally irrelevant, but never marginally relevant \cite{Green:2010da}. A space of exactly marginal operators exists, in general, and near the origin, namely around $g_i=0$, it is described by the quotient
\be
\label{confman1}
{\cal M}_c = \{ g_i | D^a = 0\}/G~~~\mbox{with}~~~ D^a= g^i T^a_{ij} \bar g^j~.
\ee
Equivalently, ${\cal M}_c = \{ g_i\}/G^\mathbbmtt{C}$, where $G^\mathbbmtt{C}$ is the complexified broken symmetry group.\footnote{For a discussion of the holographic counterpart of these results see Refs.~\cite{Tachikawa:2005tq,Louis:2016qca,Aharony:2002hx,Ashmore:2016oug}.} To summarize, the conformal manifold is parametrized by all uncharged operators (which trivially satisfy the constraint $D^a=0$ and are hence exactly marginal by themselves) plus all $G$-inequivalent linear combinations of charged, classically marginal operators ${\cal O}_i$ satisfying the constraint \eqref{confman1}.

There can exist submanifolds of ${\cal M}_c$ where only a subgroup $H \subset G$ of the global symmetries is preserved. Along such submanifolds, current multiplets belonging to the complement of $H$ in $G$ recombine. These are the submanifolds we will be interested in.


\section{Multiplet recombination along conformal manifolds}
 \label{marsec}

As discussed above, the existence of exactly marginal deformation, and in turn of a conformal manifold, is a generic property of supersymmetric field theories. Hence, in what follows, we will stick to four-dimensional $\cn=1$ SCFT. We will first present a toy model and then consider a class of models which naturally arises in string theory, namely, SCFT describing the dynamics of D3-branes at toric Calabi-Yau singularities.

\subsection{Abelian toy model}
\label{fttoy}

Let us consider a four-dimensional $\cn=1$ SCFT admitting a $U(1)$ global symmetry, and assume there exists $n$ chiral primary (classically) marginal operators ${\cal O}_i$ with charge $q_i$ under $U(1)$. A generic symmetry-breaking deformation can be described by the action
\be
S= S_{\tx{SCFT}} + \sum_i \int d^4x \, g_i O_i  + h.c.~,
\ee
where $O_i$ are the F-components of the chiral superfields ${\cal O}_i$ and $g_i$ are complex couplings.

The submanifold of ${\cal M}_c$ along which the $U(1)$ symmetry is broken is described by the D-term-like equation
\be
\label{du0}
\sum_{i,=1}^n  q_i \, g_i \bar g_i   = 0 ~,
\ee
modulo $U(1)$ transformations. There exist $n-1$ non-trivial solutions of the above equation, in general. Let us dub $O$ a linear combination of operators $O_i$ which solves eq.~\eqref{du0}
\be
O = g_1 O_1 + g_2 O_2 + \dots + g_n O_n~.
\ee
This is an exactly marginal deformation. Hence, if perturbing the original SCFT with ${\cal W} = g \, {\cal O}$, one describes yet another SCFT, which is parametrically near to the original one as $g \rightarrow 0$. In the deformed SCFT, the $U(1)$ symmetry is broken and the $U(1)$ current is not conserved
\be
\mbox{SCFT}_0:~~\partial_\mu J^\mu =0~~~,~~~\mbox{SCFT}_g:~~\partial_\mu J^\mu \not=0~.  
\ee
The minimal number of marginal operators which can provide non-trivial solutions of eq.~\eqref{du0} is two. In the following, we will then consider, for definiteness, $i=1,2$. In this case, there exists a one-dimensional subspace in the space of couplings which corresponds to an exactly marginal deformation, described by the equation
\be
\label{du1}
q_1\, |g_1|^2 + q_2 \, |g_2|^2=0~, 
\ee
modulo $U(1)$ transformations. The general solution is $g_1=\sqrt{-q_2/q_1}e^{i\f}g_2\equiv g$, with $\f$ an arbitrary phase.\footnote{Note that from eq.~\eqref{du1} it follows that $q_1$ and $q_2$ should have opposite sign.} Within this set we can choose a convenient representative. Upon a $U(1)$ rotation
\be
O_1 \rightarrow e^{i q_1 \alpha} O_1 ~~,~~O_2 \rightarrow e^{i q_2 \alpha} O_2~.
\ee
Choosing $\alpha = \f/(q_2 - q_1)$, and fixing for definiteness $q_1 = -q_2 \equiv q$ we get for the representative
\be
O_+ \equiv O_1 + O_2~,
\ee
and the symmetry-breaking SCFT is described by the action
\be
\label{defu2}
S_{\tx{SCFT}}^\prime =  S_{\tx{SCFT}} + \int d^4x \,  g  O_+ + h.c.~.
\ee
Note that, once this parametrization is chosen, any combinations of $O_1$ and $O_2$ not proportional to $O_+$ itself will be marginally irrelevant (in particular, the operator $O_- \equiv O_1 - O_2$).

By Noether method, one can compute the current (non) conservation equation, which reads
\be
\label{opi1}
\partial_\mu J^\mu = i q~ g \, O_- + h.c.~.
\ee
The fact that $O_-$ is (marginally) irrelevant nicely agrees with $\Delta_J $ being bigger than 3 whenever $g \not = 0$.
 
To leading order in $g$ the anomalous dimension of the current $J_\m$ can be computed following the approach reviewed in section \ref{secmethod}. The result is 
\be
\label{lgreltoy}
\g  =  \frac14 q^2 |g|^2 \, \frac{C_{O_-}}{C_J}  + O(g^4)~,
\ee
where $C_{O_-}$ is the normalization of the two-point function $\la O_-{O}^\dagger_- \ra$.\footnote{The discrepancy in the numerical coefficient with eq.~\eqref{lgCrel2} is because the deformation considered here is complex, compare eq.~\eqref{opi1} with eq.~\eqref{brokencur2}.} Here, $C_{O-}$ and $C_J$ are to be evaluated at $g=0$, so are data of the undeformed SCFT. 

For interacting CFTs it may happen that the coupling $\l$ governing their dynamics is itself exactly marginal and the free limit, $\l =0$, is part of the conformal manifold  (this is the case for  $\cn=4$ SYM, which we will consider later). If a holographic description is available, one could then compute eq.~\eqref{lgreltoy} for small and large values of $\l$,  and compare. In general, one should expect different answers for $\g$ at small and large $\l$.  
A simplification is that the coefficients entering eq.~\eqref{lgreltoy} are to be evaluated at $g=0$. At any $\l$, the symmetry is preserved for $g=0$ and, for a conserved current, the coefficient $C_J$ of the two-point function does not renormalize.\footnote{This is because in a SCFT the coefficient $C_J$ of the two-point function of a conserved non-R current is nothing but the cubic  't Hooft anomaly between the superconformal R current and the current ${\rm Tr}\,(T_RT_JT_J)$ itself \cite{Anselmi:1997ys,Barnes:2005bw}. As such, it does not depend on $\l$.} On the contrary, nothing like this is expected to hold for the operator $O_-$ and so for $C_{O_-}$, in principle. In fact, supersymmetry can also protect $C_{O_-}$, sometimes, as we will see later.

\subsection{$\b$ deformed superconformal field theories}
\label{bdef}

D3-branes at conical Calabi-Yau (CY) singularities, that is real cones over Sasaki-Einstein manifolds $X_5$, provide a large class of $\cn=1$ SCFT with holographic duals, the dual geometry being $AdS_5 \times X_5$. The most studied examples are toric CY, which are CY for which $X_5$ admits at least a $U(1)^3$ isometry group. Of these three abelian factors, one (that associated to the Reeb vector)  corresponds to the superconformal R symmetry. The other two are flavor symmetries of the dual field theory.

For any toric CY singularity there always exists a supersymmetric, exactly marginal deformation preserving the $U(1)^3$ symmetry \cite{Benvenuti:2005wi}. This is known as $\b$ deformation. It may happen that $X_5$ has an enlarged isometry group $H  \supset U(1)^3$. In this case, the $\b$ deformation triggers current multiplet recombination since by $\b$ deforming the theory the flavor group $H$ is broken to $U(1)_R \times U(1)^2$ and several currents are not conserved anymore.\footnote{Note that exactly marginal deformations do not break conformal symmetry and therefore always preserve the superconformal R current.}  This is the class of models of interest in our present analysis.

In what follows, we will discuss three such examples: the $\b$-deformed $\cn=4$ SYM, the $\b$-deformed conifold theory and the $\b$-deformed  $Y^{p,q}$ theories. In the first case, $H = U(1)_R \times SU(3)$.\footnote{From a $\cn=1$ perspective, the $SU(4)$ R-symmetry group of $\cn=4$ SYM should be seen as $U(1)_R \times SU(3)$, with the abelian factor being the $\cn=1$ R symmetry and $SU(3)$ a flavor symmetry.} For the conifold theory, $H = U(1)_R \times SU(2) \times SU(2)$, while for $Y^{p,q}$ singularities $H = U(1)_R \times SU(2) \times U(1)$. 

These models share many similarities, but there is one sharp difference: for $\cn=4$ the free theory is part of the conformal manifold. For the conifold and $Y^{p,q}$ theories, it is not \cite{Strassler:2005qs}. Therefore, in the latter cases the only available tool to compute current anomalous dimension is AdS/CFT. In the $\b$-deformed $\cn=4$ theory, instead, one can compute current anomalous dimensions at both weak and strong coupling. 

In preparation for what we do next, let us recall some basic results about the structure of the conformal manifold for these theories. 

\subsubsection*{On conformal manifolds of toric Calabi-Yau singularities}
\label{n4b}

The space of exactly marginal deformations of $\cn=4$ SYM is three dimensional \cite{Leigh:1995ep}. Besides the one associated to the complex gauge coupling, which preserves all flavor symmetries, there exist two $\cn=1$ preserving deformations: the $\b$ deformation, which preserves a $U(1)^2$ of the flavor-symmetry group, and the so-called cubic deformation, which breaks the flavor-symmetry group fully. We will be interested in the $\b$ deformation, which is generated by the superpotential 
\be
\label{bn4} 
\cw_\b = \l_\b \, \tx{Tr}\(\F_1 \F_2 \F_3 + \F_1 \F_3 \F_2 \)~,
\ee
where $\F_i$ are the three adjoint chiral superfields of the $\cn=4$ vector multiplet and transform in the $\bf{3}$ of $SU(3)$. 
 
The SCFT describing the dynamics of D3-branes at the tip of the conifold (a CY with $X_5 = T^{1,1}$ whose topology is $S^3 \times S^2$) \cite{Klebanov:1998hh} is a four-dimensional $\cn=1$ superconformal gauge theory with gauge group $SU(N)\times SU(N)$, a flavor-symmetry group $SU(2) \times SU(2)$, bi-fundamental matter and a quartic superpotential
\be
\label{KWdef}
\cw= \l_{KW}\, \e^{\a\b} \e^{\dot{\a}\dot{\b}} \tx{Tr}\(A_\a B_{\dot{\a}} A_\b B_{\dot{\b}}\)~,
\ee 
where $\a$ and $\dot\a$ are flavor indices, corresponding to the two $SU(2)$ factors, respectively. The fields $A_\a$ transform in the $(\frac 12 , 0)$ of the flavor-symmetry group $SU(2) \times SU(2)$. The  $B_{\dot\a}$ transform instead in the $(0, \frac 12)$. 

The conformal manifold of the conifold theory is a five-dimensional space \cite{Benvenuti:2005wi}. Two exactly marginal deformations, parametrized by suitable functions of the superpotential coupling $\l_{KW}$ and the sum and difference of the inverse gauge coupling squared \cite{Klebanov:1998hh}, are invariant under $SU(2) \times SU(2)$. The other three break the flavor-symmetry group. As already emphasized, an important difference with respect to $\cn=4$ SYM is that the free theory, $g_1=g_2=0$, is not part of the conformal manifold \cite{Strassler:2005qs}. This means that in computing eq.~\eqref{lgCrel2}, there is no regime where a field theory, perturbative analysis applies. 

Holographically, each exactly marginal deformation is associated to a massless excitation in the bulk. The dilaton and the $B_2$ flux over $S^2$ are dual to the flavor-singlet deformations. The flavor-breaking deformations are instead associated to excitations of KK modes. Of these, the $\b$ deformation, which preserves a  $U(1)^2$ flavor symmetry, corresponds to the following superpotential coupling 
\bea
\label{bcon}
\cw_{\b} &=& 
\l_\b \, \tx{Tr}\(A_1B_{\dot{1}} A_2 B_{\dot{2}} + A_1 B_{\dot{2}} A_2 B_{\dot{1}}\) ~.
\eea

The conifold theory is in fact part of an infinite class of $\cn=1$ SCFT which arises by considering D3-branes at CY singularities whose bases are the so-called $Y^{p,q}$ manifolds  \cite{Gauntlett:2004zh,Martelli:2004wu}. These are Sasaki-Einstein manifolds with the same topology of the conifold (the conifold is nothing but a real cone over $Y^{1,0}$), but with different properties for generic $p,q$; {\it e.g.}, the R charges are irrationals \cite{Bertolini:2004xf,Benvenuti:2004dy}. The flavor-symmetry group is $SU(2) \times U(1)$, there are $2p$ $SU(N)$ gauge groups and $4p+2q$ bi-fundamental fields of four different types, $U^\a, V^\a, Y$ and $Z$, with $\a$ an $SU(2)$ flavor index. The properties of these fields are summarized in appendix \ref{quantProp}. Finally, there is a superpotential with cubic and quartic couplings
\be
\label{Ypqdef}
\cw =  \sum_{i=1}^q \e_{\a\b} \, \tx{Tr} \(U^\a_i V^\b_i Y_{2i-1} + V_i^\a U^\b_{i+1} Y_{2i}\) +   \sum_{j=q+1}^p \e_{\a\b} \, \tx{Tr} \(Z_j U^\a_{j+1} Y_{2j-1} U^\b_j\)~.
\ee 
The conformal manifold is three dimensional \cite{Benvenuti:2005wi}. Two exactly marginal deformations are flavor singlets and correspond to the dilaton and the $B_2$ flux, as for the conifold. The third breaks the flavor group to $U(1)^2$ and is described by the superpotential coupling
\be
\label{bYpq}
\cw_{\b} = \l_\b \, \tx{Tr} \( \sum_{i=1}^q \sigma_{3\a}^{~\b} \(U^\a_i V_{i \b} Y_{2i+2} + V_i^\a U_{i+1 \b} Y_{2i+3}\) +   \sum_{j=q+1}^p \sigma_{3\a}^{~\b} \, Z_j U^\a_{j+1} Y_{2j+3} U_{ j \b}\) ~,
\ee
where $\sigma_3$ is a Pauli matrix. As for the conifold theory, the free theory is not part of the conformal manifold. 

By performing a $\b$ deformation in the $\cn=4$, conifold and $Y^{p,q}$ theories, several global currents acquire an anomalous dimension. Our aim will be to compute the leading correction to $\g$, eq.~ \eqref{lgCrel2}, where $g$ here is $\l_\b$ and ${\cal O}$ are chiral primaries obtained acting with a flavor-symmetry transformation on the operators \eqref{bn4}, \eqref{bcon} and \eqref{bYpq} respectively, at $\l_\b=0$. To this aim, we need to compute the two-point functions of these scalar operators (actually of their F-components) and of the corresponding broken currents at $\l_\b=0$. For the conifold and the $Y^{p,q}$ series, this is a computation inherently at strong coupling, hence the only available tool is AdS/CFT. For $\cn=4$ instead, one could evaluate the current anomalous dimension at both weak and strong coupling, since the free theory belongs to the conformal manifold in this case. However, for $\cn=4$ well-known non-renormalization theorems ensure that, as far as eq.~\eqref{lgCrel2} is concerned, the weak and strong coupling results are the same: the two-point function one has to compute involves 1/2 BPS operators, and this is known not to renormalize \cite{Lee:1998bxa}(recall we have to evaluate at  $\l_\b=0$). Therefore, in what follows we will treat all three cases holographically. 

The gravity dual of $\b$-deformed $\cn=4$ SYM and more general toric singularities, including the conifold and the $Y^{p,q}$ series, was found in Ref.~\cite{Lunin:2005jy} (see also Ref.~\cite{Butti:2007aq}). This will allow us to treat the three different cases somewhat together.  

\subsubsection*{Broken currents anomalous dimensions}\label{bcad}

In an $\cn=1$ SCFT with chiral superfields $\F_i$ the coefficient $C_J$ appearing in eq.~\eqref{2pfj}, can be computed using the R charges and flavor quantum numbers of fermions in the theory via the following 't Hooft anomaly \cite{Anselmi:1997ys}
\be
C_J= 36 \sum_i\(\dim R_i\)(1-r_i)\Tr_i\(T^a T^b\)~.
\ee
Here, $r_i$ are the R charges of the chiral superfields and $R_i$ the representations they transform under gauge-symmetry transformations (R charges of chiral superfields are reported in appendix \ref{quantProp}). The values of $C_J$ for the various theories are presented in table \ref{cjValues}. The non-abelian flavor-symmetry generators are in the fundamental representation and are normalized as $\Tr \(T^aT^b\)=\frac12 \d^{ab}$. Note that, consistently, the result for $Y^{p,q}$ theories is positive definite, hence satisfying unitarity, for $p\geq q\geq 0$, which is the range for which $Y^{p,q}$ manifolds are defined. 
\begin{table}[h]
\begin{center}
$$\begin{array}{| c | c |c |}  \hline
\rowcolor[gray]{0.8}
\mathrm{Theory} & \tx{Flavor group} & \tx{Current central charge: }C_J  \\ \hline\hline
   \cn=4~ \mathrm{ SYM} & SU(3)& 6(N^2-1) \\  \hline
   \mathrm{Conifold~theory} &SU(2)\times SU(2)& 9N^2\\  \hline
   \multirow{2}{*}{$Y^{p,q}~ \mathrm{ theories}$} & SU(2) & 6N^2\(5 p q^2-4 p^3+\left(2 p^2-q^2\right) \sqrt{4 p^2-3 q^2}\)/q^2 \\ \cline{2-3}
 & U(1) & 48N^2 p^2\left(2p-\sqrt{4 p^2-3 q^2}\right)/q^2  \\ \hline
\end{array}$$
\caption{Central charges for the non-anomalous global currents}
\label{cjValues}
\end{center}
\end{table}

Next we turn to the calculation of $C_O$. Since ${\cal O}$ and $\cw_\b$ lie in the same representation of the flavor group, up to a group theory factor (which for all cases we consider turns out to be 1) they have the same normalization. Therefore, the value of $C_O$ is the same as the value of the corresponding $C_{W_{\b}}$ (which is nothing but the component of the Zamolodchikov metric along the corresponding modulus).\footnote{Here $W_\beta$ is the F-component of the superpotential $\cw_\b$.} The two-point function for $W_\b$ can be extracted from the bulk effective action for the dual massless scalars $\beta$ which is known to be \cite{Lunin:2005jy}
\be
\label{5Daction}
S=-\frac{N^2}{16\pi^2 R_E^3}\int d^5x\sqrt{g} \[C\frac{\pa_\m\b~\pa^\m\ol{\b}}{\Im \t}~\],~~~~\b=\g-\t\s~,
\ee
where $\t$ is the axio-dilaton, $R_E$ is the radius of AdS and the normalization $C$ depends on the geometry of the compact manifold $X_5$ and reads
\be\label{Ccoefficients}
C=\la g_{0,E}\ra \frac{\tx{Vol}(S^5)}{\tx{Vol}(X_5)}~.
\ee
In the above expression $\la g_{0,E} \ra$ is the average value of the determinant of the metric on the internal 2-torus that geometrically realizes the $U(1)\times U(1)$ symmetry in the dual field theory.  The values of $\la g_{0,E}\ra$ in the three cases are presented in appendix \ref{coefficients}. The two-point function for the marginal operator $W_\b$ that one derives from \eqref{5Daction} is 
\be
\la W_\b(x) {W}^\dagger_\b(0)\ra=\frac{N^2}{(2\p)^4}\frac{C}{\Im\t}\frac{4!}{|x|^8}~,
\ee
assuming a bulk/boundary coupling with unit normalization,  $\int d^4x ~ \b \, W_\b+h.c.$.\footnote{This is suggested by the fact that both parameters are periodic with the same period \cite{Lunin:2005jy}.}
This gives the value of $C_{W_\b}$ and in turn $C_O$ 
\be
C_O=24 g_s N^2 C~.
\ee
In Table \ref{cOValues} we list the value of $C_O$ for various theories under consideration. 
\begin{table}[h]
\begin{center}
$$\begin{array}{| c | c |c |}  \hline
\rowcolor[gray]{0.8}
\mathrm{Theory} & C_O  \\ \hline\hline
   \cn=4~ \mathrm{ SYM} & 24\pi g_s N^3 \\  \hline
   \mathrm{Conifold~theory} & 45\p g_s N^3/2\\  \hline
   Y^{p,q}~ \mathrm{ theories} & 24\pi g_s N^3 p \left(7 p q^2-8 p^3+\left(4 p^2-2 q^2\right) \sqrt{4 p^2-3 q^2}\right)/q^4   \\ \hline
\end{array}$$
\caption{Normalization of two-point functions of the marginally irrelevant operators.}
\label{cOValues}
\end{center}
\end{table}

Plugging the value of $C_O$ and $C_J$ in eq. \eqref{lgCrel2} (and remembering that these deformations are complex) we obtain the values of $\g$. Table \ref{anomlDim} contains our results (to express  these anomalous dimensions in terms of the field theory parameter $\l_\b$, one should take into account that, following the conventions of Ref.~\cite{Lunin:2005jy}, there is a $(g_s)^{1/2}$ difference between $\l_\b$ and $\b$; therefore, the resulting anomalous dimensions scale just with $N$).
\begin{table}[h]
\begin{center}
$$\begin{array}{| c | c |c |}  \hline
\rowcolor[gray]{0.8}
\mathrm{Theory} & \tx{Broken Flavor group} & \tx{Current anomalous dimension: }\g  \\ \hline\hline
   \cn=4~ \mathrm{ SYM} & SU(3)& \p g_s N |\b|^2 \\  \hline
   \mathrm{Conifold~theory} &SU(2)\times SU(2)& 5\p g_s N|\b|^2/8\\  \hline
   \multirow{1}{*}{$Y^{p,q}~ \mathrm{ theories}$} & SU(2) & \p g_s N|\b|^2 \frac{p}{q^2}\(\frac{2 q^4-4 p^4+p^2 q^2+\sqrt{4 p^2-3 q^2} \left(2 p^3-p q^2\right)}{q^4-4 p^4+3 p^2 q^2}\)\\ \hline
\end{array}$$
\caption{Anomalous dimensions for the broken currents belonging to the non-Cartan elements of the flavor group.}
\label{anomlDim}
\end{center}
\end{table}

The $g_s$ and $N$ dependence of current anomalous dimensions can be equivalently obtained from the mass of the dual bulk gauge field. To see this, it is sufficient to look at the $\m-\a$ component of Einstein's equation. Schematically, we have
\be
R_{\m\a}\supset-\frac{1}{48}\frac{|G_3|^2}{\Im \t}g_{\m\a}~,
\ee
where the holomorphic 3-form flux in the $\b$ deformed geometry takes the form \cite{Lunin:2005jy}
\begin{align}\label{3form}
G_3&=-\(\g-\t \s\)R_E^4 ~d\(12 \o_1\w d\psi+i G\o_2\)~,
\end{align}
where $R_E^4=4\p N$ with $\alpha'=1$. The 2-forms $\o_1\w d\psi, \o_2$ and the function $G$ are different for different cases but the form \eqref{3form} for $G_3$ is the same for $S^5, T^{1,1} \tx{ and }Y^{p,q}$. This implies that $|G_3|^2\propto |\g-\t \s|^2 R_E^8 R_E^{-6}=|\g-\t \s|^2 R_E^2$. The extra $R_E^{-6}$ comes from the metric used for contracting the indices in $|G_3|^2$. The Maxwell operator is normalized with an additional factor of $R_E^{-2}$. Therefore, after canonically normalizing the Maxwell operator we see that the mass term is proportional to
\be
m^2\propto\frac{|\g-\t \s|^2 R_E^4}{\Im\t}=|\g-\t \s|^2 4\p N g_s\equiv 4\p g_s N |\b|^2~.
\ee
It would be interesting to reproduce the exact coefficient by analyzing the fluctuation equations for the gauge fields in detail and see whether the result matches with those in table \ref{anomlDim}. In the undeformed background, the (massless) gauge fields dual to conserved currents are degenerate and lie in the adjoint representation of the isometry group of $X_5$. When the $\b$ deformation is turned on the degeneracy is partially lifted, making some of the gauge fields (those that belong to the non-Cartan elements) massive. The $\b$ deformation turns on modes that have dependence on the $X_5$ coordinates. If the explicit form of vector spherical harmonics on $X_5$ were known, it would become possible to perform degenerate state perturbation theory and obtain the mass splitting to leading order in the deformation. In this sense, our results for the anomalous dimensions in Table \ref{anomlDim} give a prediction for bulk gauge field masses in the deformed background, to leading order in $\beta$.

\section{Multiplet recombination by relevant deformations}
\label{relsec}

In this section we consider instances of symmetry-breaking relevant deformations. First we describe symmetry breaking in the $O(N)$ model in $d=4-\epsilon$ dimensions. In the free phase, this theory admits a $O(N)$ global symmetry. One can add symmetry-breaking relevant deformations, which induce a RG flow which brings the theory to an IR fixed point of the Wilson-Fischer type where current multiplets recombine. For sufficiently small $\epsilon$, the IR fixed point is parametrically near to the UV free phase, and one can rely on pure field theory techniques to compute current anomalous dimensions. Later, we focus on a class of RG flows interpolating between $\cn=1$ SCFTs at strong coupling. These flows are described, holographically, by AdS-to-AdS BPS domain-wall solutions of a simple $\cn=2$ gauged five-dimensional supergravity model, originally discussed in Ref.~\cite{Ceresole:2001wi}.

\subsection{The $O(N)$ model}

The action of the free $O(N)$ model in $d$ dimensions is
\be 
S=\frac12\sum_{i=1}^N \int d^dx~\(\pa\f_i\)^2~,
\ee 
where $\f_i$ are $N$ real scalar fields. This theory possesses a global $O(N)$ symmetry, and the set of corresponding currents reads
\begin{equation}
\label{curform1}
J^a_{\mu}=-\partial_{\mu}\phi_i(T^a)^{ij}\phi_j,~~~~~~~~ a=1,..,\frac{N(N-1)}{2}~,
\end{equation}
where $T^a$ are generators of $O(N)$ (normalized here as $\Tr\(T^aT^b\)=-2\delta^{ab}$). Using the scalar two-point function 
\be 
\<\phi_i(x)\phi_j(0)\>=\frac{\delta_{ij}}{(2\pi)^{d/2}|x|^{d-2}}~, 
\ee
we get the following two-point function for the currents
\be 
\< J_{\mu}^a(x)J_{\nu}^b(0)\>=2(d-2)\delta^{ab}\frac{I_{\m\n}}{(2\pi)^d|x|^{2d-2}}~,
\ee
where $I_{\m\n}$ is defined in eq.~\eqref{2pfj}. In $d=4-\e$ dimensions we see that $C_{J}=4-2\e$. We would like to deform this theory via a relevant deformation such that the resulting theory has a fixed point with (partially) broken global symmetry. To this end, let us consider the following deformation which breaks $O(N)$ to $O(N-1)$
\be 
\label{defOn}
S_{\tx{def}}=\int d^dx\(\frac{g_1}{4!}\phi_1^4+\frac{g_2}{4}\phi_1^2\sum_{j=2}^N\phi_j^2+\frac{g_3}{4!}\big(\sum_{j=2}^N\phi_j^2\big)^2\)~.
\ee 
Let us first choose $g_2=0$. In this case, we have two decoupled sectors, a $\f^4$ theory and an interacting $O(N-1)$ model (which implies that $g_2$ will not be generated quantum mechanically either). 

The RG flow resulting from this deformation ends up in a weakly interacting IR fixed point of the Wilson-Fisher type where the values of the couplings $(g_{1*},g_{3*})$ are
\be\label{fixedpoints}
g_{1*}=\frac{16\pi^2}{3}\epsilon+\mathcal{O}(\epsilon^2),~~~~ g_{3*}=\frac{48\p^2}{N+7}\e +\mathcal{O}(\epsilon^2)~.
\ee
The deformation gives rise to the following anomalous currents which were otherwise conserved
\begin{equation}{\label{noncons1}}
\partial^{\mu}J^a_{\mu}=-\frac{g_{1*}}{3!}(T^a)^{1j}\phi_j\phi_1^3+\frac{g_{3*}}{3!}\(T^a\)^{1j}\sum_{k=2}^N\f_1\f_j\f_k\f_k~.
\end{equation}
In total there are $N-1$ broken currents. Computing the two-point functions of operators on the right-hand side, which provides the values of $C_O$, one finally gets from eq.~\eqref{lgCrel2}  
\begin{equation}
\gamma_J=\(\frac{1}{108}+\frac{N+1}{4(N+7)^2}\)\epsilon^2+\mathcal{O}(\epsilon^3)~.
\end{equation}
As shown in appendix \ref{betaFunc}, this is nothing but the sum of the anomalous dimensions of constituent fields, $\f_1$ and $\f_j$ ($j\ne 1$). This is expected because for $g_2=0$  the broken currents are composed of fields belonging to decoupled sectors.

The symmetry-breaking pattern we discussed here is an instance of the more general one $O(N) \to O(N-M) \times O(M)$, which can be obtained  by a straightforward generalization of the action \eqref{defOn}
\begin{equation}
\label{defMN}
\int d^dx\Bigg[\frac{g_1}{4!}\(\sum_{i=1}^M\phi_i^2\)^2+\frac{g_2}{4}\sum_{i=1}^M\phi_i^2\sum_{j=M+1}^N\phi_j^2+\frac{g_3}{4!}\(\sum_{j=M+1}^N\phi_j^2\)^2\Bigg]~.
\end{equation}
Again, for $g_2=0$ there are two decoupled sectors and current anomalous dimensions are given by the sum of elementary fields anomalous dimensions. The computations one needs to do are basically the same we did before, and we refrain from presenting them here. 

By computing derivatives of the $\b$ functions one can check the stability of these fixed points. It turns out that some of them are unstable (they are saddle points), meaning that they could be reached only by some fine-tuning of the UV couplings. Note, however, that, regardless of their nature, current multiplet recombination occurs at any of these fixed points, and one can compute anomalous dimensions of broken currents through the method we described in section \ref{secmethod}.

Let us now consider a deformation with $g_2\ne 0$. Regardless of their tree-level values, quantum mechanically also $g_1$ and $g_3$ are generated now, so in this case one has to confront with the most general deformation. Looking at the $\b$ functions of the three couplings, which we report in appendix \ref{betaFunc}, one sees that for specific values of $M$ and $N$ there exist new fixed points having $g_{2*} \not = 0$. This implies that elementary field sectors constituting the broken currents are no longer decoupled and, in turn, that current anomalous dimensions are not just the sum of those of elementary fields. A simple case to look at is $N=2M$. One finds that for any $M$ a fixed point exists where 
\begin{align}
&g_{1*}=\frac{24 \pi ^2 \epsilon M}{M^2+8}~,~~~~g_{2*}= \frac{8 \pi ^2 \epsilon (4-M) }{M^2+8}~,~~~~g_{3*}= \frac{24 \pi ^2 \epsilon M}{M^2+8}~,
\end{align}
with current anomalous dimension 
\begin{equation}
\gamma_J=\frac{\epsilon^2 (M+2)(M-2)^2}{2(M^2+8)^2}~.
\end{equation}
More details can be found in Appendix \ref{betaFunc}.

\subsection{AdS-to-AdS domain walls}
\label{longrg}

We would like now to consider symmetry-breaking relevant deformations connecting $\cn=1$ SCFT at strong coupling. This is outside the realm of (perturbative) QFT, and hence we will rely on holography. Flows of this kind are described by BPS solutions of five-dimensional $\cn=2$ supergravity with an AdS-to-AdS domain-wall metric and one or more scalars having non-trivial profiles. 

Note that in five-dimensional $\cn=2$ supergravity scalars belong either to hypermultiplets or to vector multiplets. The former are dual to chiral operators, the latter to real linear multiplets  (which contain the spin-one currents). Therefore, flows triggered by superpotential deformations imply that hypermultiplet scalars in general run. If the chiral operators are charged under a given symmetry, the corresponding bulk gauge fields undergo a Higgs mechanism and so, by supersymmetry, also the vector multiplet scalars are expected to run. 

As an illustrative example, we consider below one such scenario. This corresponds to a SCFT with $U(1)_{\tilde R} \times U(1)$ symmetry (the always-present superconformal R symmetry and an abelian flavor symmetry) perturbed by a charged, relevant deformation ${\cal O}$ triggering a RG flow towards an IR fixed point. If there are no emergent symmetries in the IR, at such a fixed point only a $U(1)_R$ superconformal R-symmetry is preserved.\footnote{The IR R symmetry is different from the UV one; i.e., it is a combination of the original R symmetry and the (broken) flavor symmetry. Indeed, a relevant deformation breaks explicitly conformal invariance and in turn the superconformal UV R symmetry $U(1)_{\tilde R}$.} The current associated to the $U(1)$  symmetry recombines and acquires an anomalous dimension.   

A two-parameter family of $\cn =2$ supergravity theories describing flows of this kind was derived long ago \cite{Ceresole:2001wi}. This is $\cn=2$ supergravity coupled to a vector multiplet and a hypermultiplet, with scalar manifold
\begin{equation}
\label{manifold1}
\mathcal{M} = O(1,1)\times \frac{SU(2,1)}{SU(2)\times U(1)}~.
\end{equation}
The first factor is parametrized by the  vector multiplet real scalar $\rho$, while the second factor by the four scalars belonging to the hypermultiplet, $q^X = (V, \sigma, \theta, \tau)$. The two gauge fields, the graviphoton $A_M$ and the one sitting in the vector multiplet, $B_M$, gauge a $U(1) \times U(1)$ subgroup of the isometry group of the hyperscalar manifold. The graviphoton is dual to the R symmetry, and the gauge field $B_M$ to the $U(1)$ flavor symmetry.

This theory admits different classes of solutions, depending on the gauging. For instance, there exist (a) domain-wall solutions which provide a holographic version  \cite{Bertolini:2013vka} of the so-called $\tau_U$ conjecture, originally proposed in Ref.~\cite{Buican:2011ty}, (b) non-supersymmetric solutions which have been used to construct models of (holographic) gauge mediation \cite{Argurio:2012cd}. We will focus, instead, on supersymmetric AdS-to-AdS solutions.

This model has been widely studied and we refer to Ref.~\cite{Ceresole:2001wi} for any technical detail. In what follows we just summarize the results we need for our analysis. 

What we are interested in are supersymmetric solutions admitting a critical point (i.e. an AdS stationary point of the gravity superpotential) which preserves a $U(1) \times U(1)$ symmetry, and a second critical point preserving a $U(1)$ symmetry. As discussed in Ref.~\cite{Ceresole:2001wi} (see also  Ref.~\cite{Bertolini:2013vka}), the existence of such fixed points selects a subclass of gaugings, parametrized by two real parameters, $\beta$ and $\g$, subject to the condition 
\begin{equation}
\label{conbz1}
(\beta-1)(1-2\zeta)>0\quad \cap\quad \zeta>0~~~\mbox{where}~~~\zeta = \frac{1-\b}{2\g -1}~.
\end{equation}
The UV and IR fixed points sit at
\bea
P_{UV} &:& q^X = (1,0,0,0) ~,~\rho=1 \\
P_{IR} &:& q^X = (1 - \xi^2,0, \xi \cos \varphi, \xi \sin \varphi) ~,~\rho= (2 \zeta)^{1/6}~,
\eea
in field space, with
\be
\xi =\sqrt{\frac{2-4\zeta}{3\beta-1-4\zeta}}~~,~~ \varphi\in \left[0,\,2\pi\right]~.
\ee
Note that $P_{IR}$ is in fact a circle of stationary points, parametrized by $\varphi$. This is an exactly marginal deformation of the IR SCFT, which does not play any role for what we want to do next. 

For any value of $\beta$ and $\gamma$ satisfying the constraint \eqref{conbz1}, there exists a smooth domain-wall (numerical) solution interpolating between $P_{UV}$ and $P_{IR}$  \cite{Ceresole:2001wi,Bertolini:2013vka}. Since $P_{UV}$ and $P_{IR}$ preserve different symmetries, these domain walls describe, holographically,  RG flows along which current multiplets recombine. Note that, as advertised, both the hyperscalars and the real scalar $\rho$ run (they have different values at $P_{UV}$ and $P_{IR}$). 

To read the gauge field masses, the relevant part of the $\cn=2$ Lagrangian is
\be
\label{n2Am}
-\frac14 a_{IJ}F^I_{\m\n}F^{J\m\n}-\frac12 \(g^2 g_{XY} K_I^X K_J^Y\) A_\m^I A^{\m I}~,
\ee
where $a_{IJ}$ is a function of the vector scalar multiplet $\rho$, $g$ controls the value of the cosmological constant and $g_{XY}$ is the metric on the hyperscalar manifold. The Killing vectors are functions of the scalar fields, hence the gauge symmetry can be Higgsed or exactly realized depending on the scalar profiles. All flows interpolating between $P_{UV}$ and $P_{IR}$ admit a vanishing Killing vector \cite{Ceresole:2001wi}, hence a massless gauge field and, correspondingly, a preserved $U(1)$ symmetry (which can be shown to be an R symmetry \cite{Ceresole:2001wi,Bertolini:2013vka}). This reduces to the superconformal R symmetry $U(1)_{\tilde R}$ in the UV and to the superconformal R-symmetry $U(1)_{R}$ in the IR. The second Killing vector, associated to the gauge field $B_M$,  instead, vanishes at $P_{UV}$, only. This implies that $B_M$ is massless at the UV fixed point, and massive elsewhere. Evaluating \eqref{n2Am} on the IR endpoint of the flow, one finds, in units of the IR AdS radius $L_{IR} =(g \, W_{IR})^{-1}$ (where $W_{IR}$ is the value of the supergravity superpotential at $P_{IR}$)
\begin{align}
m_A^2&=0~~,~~
\label{mbir}
m_B^2=  \frac34 \(\frac{(2 \beta +2 \gamma -3) (6 \beta  \gamma +\beta -2 \gamma -3)}{ (2 \gamma -1)^{4/3} \left(1-\beta\right)^{2/3}}\)~.
\end{align}
Plugging the above formula into the mass/dimension relation \eqref{massdimcur1} one gets the holographic prediction for the $U(1)$ flavor current anomalous dimension. 

As a consistency check, one can evaluate \eqref{mbir} for $\b=-1, \g=\frac32$, which, as shown in Ref.~\cite{Ceresole:2001wi}, corresponds to the FGPW flow \cite{Freedman:1999gp}. This is known to describe, holographically, the $\cn=1^*$ mass deformation of $\cn=4$ theory. One gets  $m^2_B =6$ and in turn $\D=2+\sqrt{7}$, in agreement with expectations \cite{Freedman:1999gp,Ceresole:2001wi}.

The supergravity model we have considered is a prototype of more general ones. It is amusing  to see how holography lets one have control on how multiplets recombine even in RG flows which might be extremely intricate from a field theory perspective, and how it makes so the description of in principle very complicated UV/IR operator maps so transparent.

\section{Conclusions}
\label{consec}

Current multiplet recombination puts severe constraints on CFT parameters. For example, we have seen that for marginal deformations anomalous dimensions of weakly broken currents are fixed, to leading order, by the Zamolodchikov metric on the conformal manifold and by a global current central charge in the undeformed CFT. 

We have considered deformations triggered by marginal as well as relevant deformations and shown that in all cases one can compute the anomalous dimension of broken currents. For theories with a holographic dual description, one can also have control on symmetry-breaking flows which are not parametrically short in the space of coupling, and anomalous dimensions can hence be large. 

The techniques we have used can be applied to several other examples. Besides the $\b$ deformation, the conformal manifold of $\cn=4$ SYM admits another symmetry-breaking deformation, which breaks the flavor-symmetry group fully, and which can be investigated field theoretically using perturbation theory. Note, also, that at generic points of the conformal manifold of $\cn=4$ SYM supersymmetry is (partially) broken. The corresponding supersymmetry current operators acquire anomalous dimensions which one could also compute. Also, the conifold theory, besides the $\b$ deformation, admits two other exactly marginal deformations with different symmetry-breaking patterns.

We focused our attention on four-dimensional theories but there exist marginal deformations for, {\it e.g.} theories in three dimensions. An example is the $\b$ deformation of the $\cn=6$ ABJM theory \cite{Aharony:2008ug} which breaks the $SU(2) \times SU(2)$ flavor symmetry down to $U(1)^2$ \cite{Bianchi:2010cx} (and here, too, supersymmetry is partially broken)

In section \ref{longrg} we have discussed one instance where the breaking is not weak, but there exist many others which can be treated in a similar manner. Here, too, the most interesting direction would be to look for non-supersymmetric flows, or flows along which supersymmetry gets partially or even fully broken. 

\section*{Acknowledgements}

We are grateful to Lorenzo Di Pietro for bringing this problem  to our attention and for useful comments. We would also like to thank Riccardo Argurio, Oleg Lunin, Ioannis Papadimitriou, Flavio Porri and Marco Serone for helpful discussions and email exchange. We thank Riccardo Argurio and Lorenzo Di Pietro also for their feedback on a preliminary draft version. We acknowledge support from the MPNS-COST Action MP1210 ``The String Theory Universe".

\appendix

\section{$\b$ deformations: matter fields quantum numbers}
\label{quantProp}
In this appendix we report the quantum numbers of matter fields of the $\cn=4$ SYM, conifold theory and $Y^{p,q}$ theories: 
\ben
\item{$\cn=4$ SYM:}
When written in the $\cn=1$ language, $\cn=4$ SYM theory contains three chiral superfields $\F^i$ that transform in the fundamental representation of $SU(3)$. The R charges of each of these is 2/3 as is evident from the $\cn=4$ superpotential. 

\item{Conifold theory:}
The theory contains two kinds of bi-fundamental matter fields $A_\a, B_{\dot\a}$. They share the same R charge $R=1/2$, and, correspondingly, the same scaling dimension $\Delta = 3/4$. The fields $A_\a$ transform in the $(\frac 12 , 0)$ of the flavor-symmetry group $SU(2) \times SU(2)$. The  $B_{\dot\a}$ transform instead in the $(0, \frac 12)$. 

\item{$\cn=1$ $Y^{p,q}$ theories:}
The theory contains four different kinds of bi-fundamental matter fields which are either singlets or doublets under the $SU(2)$ flavor symmetry. There are $p$ doublets labelled $U_\a$, $q$ doublets labelled $V_\a$, $p-q$ singlets labelled $Z$ and $p+q$ singlets labelled $Y$. Under the $U(1)$ flavor (non-R) symmetry these fields have charges $0,1,-1,1$, respectively, whereas under the $U(1)$ R symmetry they have the following charges
\begin{align}
r_U&=\frac23 pq^{-2}\(2p-\sqrt{4p^2-3q^2}\),\NO\\
r_V&=\frac13 q^{-1}\(3q-2p+\sqrt{4p^2-3q^2}\),\NO\\
r_Z&=\frac13 q^{-2}\(-4p^2+3q^2+2pq+(2p-q)\sqrt{4p^2-3q^2}\),\NO\\
r_Y&=\frac13 q^{-2}\(-4p^2+3q^2-2pq+(2p-q)\sqrt{4p^2-3q^2}\)~.
\end{align}
\een

\section{Volumes of $X_5$ and the 2-torus}
\label{coefficients}
In this appendix we give the expressions for $\tx{Vol}{(S^5)}/\tx{Vol}(X_5)$ and $\la g_{0,E}\ra$ which were needed to derive $C_O$ in section \ref{bcad}. The ratios of the volumes defined in eq.~\eqref{Ccoefficients} are (see Ref.~\cite{Lunin:2005jy} and references therein for details)
\begin{align}
\frac{\tx{Vol}{(S^5)}}{\tx{Vol}(T^{1,1})}= \frac{27}{16}~,~~~~~\frac{\tx{Vol}{(S^5)}}{\tx{Vol}(Y^{p,q})}=\frac{3p^2\(3q^2-2p^2+p\sqrt{4p^2-3q^2}\)}{q^2\(2p+\sqrt{4p^2-3q^2}\)}~.
\end{align}
The average value of the determinant of the internal 2-torus $\la g_{0,E}\ra$ can be computed from the corresponding metrics given in Ref.~\cite{Lunin:2005jy}. We summarize them below.
\ben
\item $\bf S^5$: The 2-torus in eq.~(3.12) of Ref.~\cite{Lunin:2005jy} is parametrized by the coordinates $(\vf_1,\vf_2)$. The average volume is $\la g_{0,E}\ra=\p N$
\item $\bf T^{1,1}$: This case is slightly subtle. The 2-torus in this case is parametrized by the coordinates $\vf_{1,2}=\frac{\f_1\pm\f_2}{2}$, where $\f_{1,2}$ are the coordinates appearing in the standard line element (eq.~(A.18) of Ref.~\cite{Lunin:2005jy}) of $T^{1,1}$. Taking this into account one finds\footnote{The last equality of eq.~(4.6) in Ref.~\cite{Lunin:2005jy} has a typo. We thank O. Lunin for a discussion on this point.}  $\la g_{0,E}\ra=\frac{5\p}{9} N$.
\item $\bf Y^{p,q}$: Here the two-torus in eq (A.24) of Ref.~\cite{Lunin:2005jy} is parametrized by $(\a,\f)$. We have $\la g_{0,E}\ra =\la g_0 \ra R_E^4$, where the determinant $g_0$ and the AdS radius $R_E$ have been defined in appendix A.2 of \cite{Lunin:2005jy}. Upon computing the average we find
\be
\la g_{0,E}\ra = \frac{7p^2-6q^2-p\sqrt{4p^2-3q^2}}{9p(p^2-q^2)}\p N~.
\ee
In computing $\la g_0\ra$, we have used the relation 
\be
a=\frac12 -\frac{p^2-3q^2}{4p^3}\sqrt{4p^2-3q^2}~
\ee
for $a$,  
and the integration over the $y$ coordinate is in the range $(y_1,y_2)$
\begin{align}
y_1&=\frac{1}{4p}\(2p-3q-\sqrt{4p^2-3q^2}\),~~~~y_2=\frac{1}{4p}\(2p+3q-\sqrt{4p^2-3q^2}\)~.
\end{align}
\een

\section{Fixed points of the deformed $O(N)$ model}
\label{betaFunc}
The most general deformation of the $O(N)$ model that breaks the $O(N)$ symmetry to $O(M)\times O(N-M)$ is
\begin{equation}
\int d^dx\Bigg[\frac{g_1}{4!}\(\sum_{i=1}^M\phi_i^2\)^2+\frac{g_2}{4}\sum_{i=1}^M\phi_i^2\sum_{j=M+1}^N\phi_j^2+\frac{g_3}{4!}\(\sum_{j=M+1}^N\phi_j^2\)^2\Bigg]~.
\end{equation}
In this case the anomalous dimension of broken currents is given by
\begin{equation}
\gamma_J=\frac{1}{(4\pi)^4}\((M+2)\(\frac{g_1}{3!}-\frac{g_2}{2}\)^2+(N-M+2)\(\frac{g_3}{3!}-\frac{g_2}{2}\)^2\)~,
\end{equation}
while that of elementary fields is given by
\begin{align}
\gamma_{\phi_i}&=\frac{1}{(4\pi)^4}\((M+2)\(\frac{g_1}{3!}\)^2+(N-M)\(\frac{g_2}{2}\)^2\),\   \ i=1,\ ...,\ M \\
\gamma_{\phi_i}&=\frac{1}{(4\pi)^4}\((N-M+2)\(\frac{g_3}{3!}\)^2+M\(\frac{g_2}{2}\)^2\),\   \ i=M+1,\ ...,\ N~.
\end{align}
In agreement with general expectations, from above equations and eq.~\eqref{curform1}, it follows that whenever $g_2=0$ the anomalous dimension of broken currents equals the sum of anomalous dimensions of constituents elementary fields, but it does not otherwise.

The one-loop $\b$ functions of the couplings $g_i$ in $d=4-\e$ dimensions read\footnote{We are grateful to Andreas Stergiou and Hugh Osborn for pointing out a missing term in the $\b$ function for the $g_2$ coupling in a previous version of the paper. Accordingly, also the subsequent analysis on the fixed points has been modified.}
\begin{align}
\b_{g_1}&= -g_1 \e +\frac{1}{16\p^2}\(\frac{g_1^2}{3}(M+8)+3g_2^2(N-M)\),\NO\\
\b_{g_2}&= -g_2 \e +\frac{g_2}{48\p^2}\(g_1(M+2)+g_3(N-M+2)+12g_2\),\label{betaf}\\
\b_{g_3}&= -g_3 \e +\frac{1}{16\p^2}\(\frac{g_3^2}{3}(N-M+8)+3g_2^2 M\)~.\NO
\end{align}
Besides the fixed points in the decoupled theory, for specific values of $M$ and $N$ there exist others also when $g_2 \not =0$. Again, as for the $g_2=0$ case, not all these fixed points are stable, and some fine-tuning of tree-level couplings may be needed. Here, we present few of the many possible fixed points and specify their nature, i.e., whether they are stable or unstable. 

\ben
\item ${\bf M=1}$: This case was discussed in the main text with the coupling $g_2$ switched off. For certain values of $N$ one can also find fixed points with $g_2\neq 0$, and with broken $O(N)$ symmetry. For example, this kind of (unstable in this case) fixed point exists for $N=3$, but does not exist for $N=4$.
\item ${\bf N=2M}$: For this case, for any $M$ there exist the following zeros of the $\b$ function equations (apart from the fixed points with two decoupled sectors, when $g_2=0$):
\begin{align}
&g_{1*}= \frac{24 \pi ^2 \epsilon }{M+4} ~,~~g_{2*}= \frac{8 \pi ^2 \epsilon }{M+4} ~,~~g_{3*}= \frac{24 \pi ^2 \epsilon }{M+4} ~~~~~~~~~~~~~~~~\tx{(Stable for $M<2$)}\\
&g_{1*}=\frac{24 \pi ^2 M \epsilon }{M^2+8} ~,~~g_{2*}= \frac{8 \pi ^2 (4-M) \epsilon }{M^2+8} ~,~~g_{3*}= \frac{24 \pi ^2 \epsilon M}{M^2+8} ~~~~\tx{(Stable for $M=3$)}
\end{align}
In the first case the full $O(N=2M)$ symmetry is preserved. In the second case, which preserves $O(M) \times O(M)$, the current anomalous dimension reads
\begin{equation}
\gamma_J=\frac{\epsilon^2 (M+2)(M-2)^2}{2(M^2+8)^2}~.
\end{equation}
\een


\end{document}